**AASCIT** American Association for Science and Technology

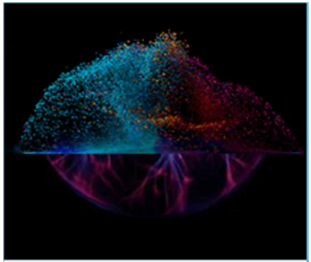

# Relativistic Modeling of Quark Stars with Tolman IV Type Potential

## Manuel Malaver

Universidad Marítima del Caribe, Departamento de Ciencias Básicas, Catia la Mar, Venezuela

### Email address
mmf.umc@gmail.com

### Citation
Manuel Malaver. Relativistic Modeling of Quark Stars with Tolman IV Type Potential.
*International Journal of Modern Physics and Application.* Vol. 2, No. 1, 2015, pp. 1-6.

### Abstract
In this paper, we studied the behavior of relativistic objects with anisotropic matter distribution considering Tolman IV form for the gravitational potential Z. The equation of state presents a quadratic relation between the energy density and the radial pressure. New exact solutions of the Einstein-Maxwell system are generated. A physical analysis of electromagnetic field indicates that is regular in the origin and well behaved. We show as the presence of an electrical field modifies the energy density, the radial pressure and the mass of the stellar object and generates a singular charge density.

### Keywords
Relativistic Objects,
Electric Field,
Gravitational Potential,
Tolman IV Type Potential,
Einstein-Maxwell System,
Charge Density

## 1. Introduction

From the development of Einstein´s theory of general relativity, the modelling of superdense mater configurations is an interesting research area [1,2]. In the last decades, such models allow explain the behavior of massive objects as neutron stars, quasars, pulsars, black holes and white dwarfs [3,4,5].

In theoretical works of realistic stellar models, is important include the pressure anisotropy [6-8]. Bowers and Liang [6] extensively discuss the effect of pressure anisotropy in general relativity. The existence of anisotropy within a star can be explained by the presence of a solid core, phase transitions, a type III super fluid, a pion condensation [9] or another physical phenomena as the presence of an electrical field [10].The physics of ultrahigh densities is not well understood and many of the strange stars studies have been performed within the framework of the MIT bag model [11]. In this model, the strange matter equation of state has a simple linear form given by $p = \frac{1}{3}(\rho - 4B)$ where $\rho$ is the energy density, p is the isotropic pressure and B is the bag constant. Many researchers have used a great variety of mathematical techniques to try to obtain exact solutions for quark stars within the framework of MIT bag model, since it has been demonstrated by Komathiraj and Maharaj [11], Malaver [12,13], Thirukkanesh and Maharaj [14], Maharaj et al. [15], Thirukkanesh and Ragel [16] and Sunzu et al. [17].

With the use of Einstein´s field equations, important advances has been made to model the interior of a star. In particular, Feroze and Siddiqui [18] and Malaver [19] consider a quadratic equation of state for the matter distribution and specify particular forms for the gravitational potential and electric field intensity. Mafa Takisa and Maharaj [20] obtained new exact solutions to the Einstein-Maxwell system of equations with a polytropic equation of state. Thirukkanesh and Ragel [21] have obtained particular models of anisotropic fluids with polytropic equation of state which are consistent with the reported experimental observations. More recently, Malaver [22,23] generated new exact solutions to the Einstein-Maxwell system considering Van der Waals modified equation of state with and without polytropical exponent and Thirukkanesh and Ragel [24] presented a anisotropic strange quark matter model by imposing a linear barotropic equation of state with Tolman IV form for the gravitational potential. Mak and Harko [25] found a relativistic model of strange quark star with the suppositions of spherical symmetry and conformal Killing vector.

The objective of this paper is to obtain new exact solutions to the Maxwell-Einstein system for anisotropic matter with an equation of state that presents a quadratic relation between the energy density and the radial pressure in static spherically symmetric spacetime using Tolman IV form for the gravitational potential Z. We have obtained some new classes of static spherically symmetrical models where the presence of an electrical field modifies the radial pressure, charge density and the mass of the compact objects. This article is organized as follows, in Section 2, we present Einstein´s field equations. In Section 3, we make a particular choice of gravitational potential $Z(x)$ that allows solving the field equations and we have obtained new models for charged anisotropic matter. In Section 4, a physical analysis of the new solutions is performed. Finally in Section 5, we conclude.

## 2. Einstein Field Equations of Anisotropic Fluid Distribution

We consider a spherically symmetric, static and homogeneous and anisotropic spacetime in Schwarzschild coordinates given by

$$ds^2 = -e^{2\nu(r)}dt^2 + e^{2\lambda(r)}dr^2 + r^2(d\theta^2 + \sin^2\theta d\varphi^2 )   \quad (1)$$

where $\nu(r)$ and $\lambda(r)$ are two arbitrary functions.

The Einstein field equations for the charged anisotropic matter are given by

$$T_{00} = -\rho - \frac{1}{2}E^2 \quad (2)$$

$$T_{11} = p_r - \frac{1}{2}E^2 \quad (3)$$

$$T_{22} = T_{33} = p_t + \frac{1}{2}E^2 \quad (4)$$

where $\rho$ is the energy density, $p_r$ is the radial pressure, $E$ is electric field intensity and $p_t$ is the tangential pressure, respectively. Using the transformations, $x = cr^2$, $Z(x) = e^{-2\lambda(r)}$ and $A^2y^2(x) = e^{2\nu(r)}$ with arbitrary constants A and c>0, suggested by Durgapal and Bannerji [26], the metric (1) take the form

$$ds^2 = -A^2y^2(x)dt^2 + \frac{1}{4cxz}dx^2 + \frac{x}{c}(d\theta^2 + \sin^2\theta d\varphi^2 )  \quad (5)$$

and the Einstein field equations can be written as

$$\frac{1-Z}{x} - 2\dot{Z} = \frac{\rho}{c} + \frac{E^2}{2c} \quad (6)$$

$$4Z\frac{\dot{y}}{y} - \frac{1-Z}{x} = \frac{p_r}{c} - \frac{E^2}{2c} \quad (7)$$

$$4xZ\frac{\ddot{y}}{y} + (4Z+2x\dot{Z})\frac{\dot{y}}{y} + \dot{Z} = \frac{p_t}{c} + \frac{E^2}{2c} \quad (8)$$

$$\sigma^2 = \frac{4cZ}{x}\left(x\dot{E} + E\right)^2 \quad (9)$$

$\sigma$ is the charge density and dots denote differentiation with respect to x. With the transformations of [26], the mass within a radius r of the sphere take the form

$$M(x) = \frac{1}{4c^{3/2}}\int_0^x \sqrt{x}\rho(x)dx \quad (10)$$

In this paper, we assume the following equation of state

$$p_r = \alpha\rho^2 \quad (11)$$

Here $\alpha$ is arbitrary constant.

## 3. A New Class of Solutions

Following Tolman [27] and Thirukkanesh and Ragel [24], we take the form of the gravitational potential, $Z(x)$ as

$$Z(x) = \frac{(1+ax)(1-bx)}{(1+2ax)} \quad (12)$$

where a and b are real constants. This potential is regular at the origin and well behaved in the interior of the sphere. We have considered the particular cases for $E^2 = 0$ and $E^2 \neq 0$.

Case I: For $E^2 = 0$ , using $Z(x)$ in eq.(2) we obtain

$$\rho = c\frac{\left[3(a+b)+\left(7ab+2a^2\right)x+6a^2bx^2\right]}{2(1+ax)^2} \quad (13)$$

Substituting (13) in eq.(11), the radial pressure can be written in the form

$$P_r = \alpha c^2\frac{\left[3(a+b)+\left(7ab+2a^2\right)x+6a^2bx^2\right]^2}{4(1+2ax)^4} \quad (14)$$

Using (13) in (10), the expression of the mass function is

$$M(x) = \frac{x^{3/2}(a+b+abx)}{2\sqrt{c}(1+2ax)} \quad (15)$$

The equations (13) and (15) have been deduced for Thirukkanesh and Ragel [24] with a linear barotropic equation of state.

The tangential pressure is given for

$$P_t = \frac{4xc(1+ax)(1-bx)}{(1+2ax)}\frac{\ddot{y}}{y} + 2c\left[\frac{2+(5a-3b)x+4a(a-2b)x^2-6a^2bx^3}{(1+2ax)^2}\right]\frac{\dot{y}}{y}$$
$$-c\frac{(a+b+2abx+2a^2bx^2)}{(1+2ax)^2} \quad (16)$$

Substituting (14) and (12) in (7), we have

$$\frac{\dot{y}}{y} = \frac{(a+b+abx)}{4(1+ax)(1-bx)}$$
$$+ \frac{\alpha c\left[3(a+b)+7abx+2a^2x+6a^2bx^2\right]^2}{4(1+2ax)^3(1+ax)(1-bx)} \quad (17)$$

Integrating (17), we obtain

$$y(x) = c_1(-1+bx)^A(1+ax)^B(1+2ax)^C \exp[D(x)] \quad (18)$$

where

$$A = -\frac{9b^3\alpha c + 16\alpha ca^2b + 24\alpha cab^2 + 4ab + 4a^2 + b^2}{4(a+b)(b+2a)},$$

$$B = -\frac{\alpha ca^2 + 4\alpha cab - a + 4\alpha cb^2}{4(a+b)},$$

$$C = \frac{17\alpha cb^2 + 4\alpha ca^2 + 28\alpha bc}{8(b+2a)}$$

and

$$D(x) = -\frac{2\alpha ac + 3\alpha bc + 4\alpha abcx}{4(1+2ax)^2} \quad (19)$$

The metric functions $e^{2\lambda}$ and $e^{2\nu}$ can be written as

$$e^{2\lambda} = \frac{(1+2ax)}{(1+ax)(1-bx)} \quad (20)$$

$$e^{2\nu} = A^2c_1^2(-1+bx)^{2A}(1+ax)^{2B}(1+2ax)^{2C}\exp[2D(x)] \quad (21)$$

Figures 1, 2 and 3 represent the graphs of $p_r$, $v_{sr}^2$, $M(x)$ with eq. (11) and $E^2 = 0$. $v_{sr}^2$ and $M(x)$ are the radial speed of sound and mass function, respectively. The graphs has been plotted for a particular choice of parameters a = 0.0169, b = 0.00454, α=1/3 with a stellar radius of r=10 km presented for Thirukkanesh and Ragel [24]. The metric for this model is

$$ds^2 = -A^2c_1^2(-1+bx)^{2A}(1+ax)^{2B}(1+2ax)^{2C}\exp[2D(x)]dt^2$$
$$+ \frac{(1+2ax)}{4xc(1+ax)(1-bx)}dx^2 + \frac{x}{c}(d\theta^2 + \sin^2\theta d\phi^2) \quad (22)$$

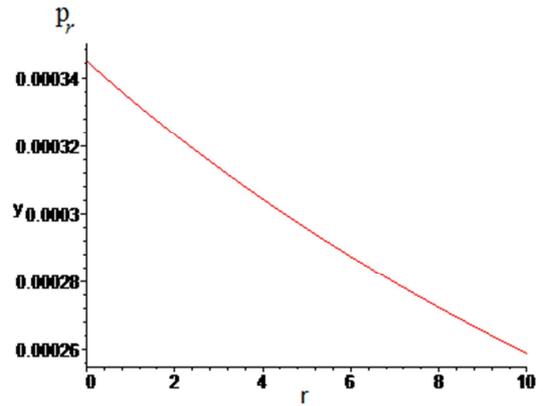

**Fig 1.** *Radial Pressure*

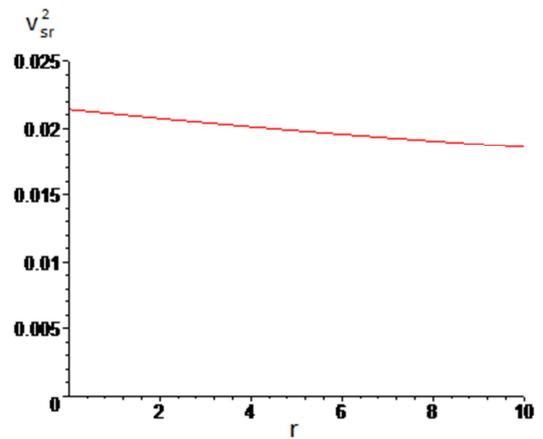

**Fig 2.** *Radial speed of sound*

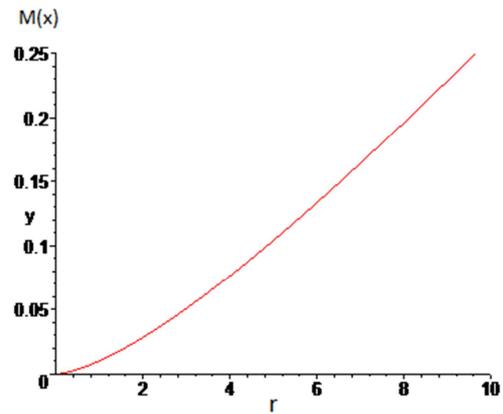

**Fig 3.** *Mass.*

Case II: For $E^2 \neq 0$ we have considered the form of the electrical field proposed for Feroze and Siddiqui [28]

$$E^2 = \frac{2c(1-Z)}{x} = \frac{2c(a+b+abx)}{1+2ax} \quad (23)$$

With eq.(23), we have found the following expressions for $\rho, P_r, M(x), \sigma^2$ and $P_t$

$$\rho = 2c \frac{\left(a + b + 2abx + 2a^2bx^2\right)}{\left(1 + ax\right)^2} \tag{24}$$

$$P_r = 4\alpha c^2 \frac{\left(a + b + 2abx + 2a^2bx^2\right)^2}{4\left(1 + ax\right)^4} \tag{25}$$

$$M(x) = \frac{\left(16a^3bx^2 + 8a^2bx - 12a^2 - 6ab\right)\sqrt{ax} + \left[\left(12a^3 + 6a^2b\right)x + 6a^2 + 3ab\right]\sqrt{2}\,\text{arctag}\sqrt{2ax}}{48a^2\sqrt{ac}\left(1 + 2ax\right)} \tag{26}$$

$$\sigma^2 = \frac{2c^2\left(1 + ax\right)\left(1 - bx\right)\left[4a^2bx^2 + a\left(2a + 5b\right)x + 2\left(a + b\right)\right]^2}{x\left(1 + 2ax\right)^4\left(a + b + abx\right)} \tag{27}$$

$$P_t = \frac{4xc\left(1 + ax\right)\left(1 - bx\right)}{\left(1 + 2ax\right)}\frac{\ddot{y}}{y}$$

$$+ 2c\left[\frac{4 + 2\left(5a - 3b\right)x + 8a\left(a - 2b\right)x^2 - 12a^2bx^3}{\left(1 + 2ax\right)^2}\right]\frac{\dot{y}}{y} \tag{28}$$

$$- c\frac{\left[2a + 2b + a\left(2a + 5b\right)x + 4a^2bx^2\right]}{\left(1 + 2ax\right)^2}$$

Substituting (23) and (25) in (7), we have

$$\frac{\dot{y}}{y} = \frac{\alpha c\left(a + b + 2abx + 2a^2bx^2\right)^2}{\left(1 + 2ax\right)^3\left(1 + ax\right)\left(1 - bx\right)} \tag{29}$$

Integrating (29), we obtain

$$y(x) = c_2\left(-1 + bx\right)^E\left(1 + ax\right)^F\left(1 + 2ax\right)^G \exp[H(x)] \tag{30}$$

Again for convenience we have let

$$E = -\frac{\alpha\left(a + b\right)cb}{2a + b}, \quad F = -\alpha c\left(a + b\right),$$

$$G = \frac{\left(6ab + 4a^2 + 3b^2\right)c\alpha}{2a + b}$$

and

$$H\left(x\right) = \frac{\left(8a^2x + 2a - b\right)c\alpha}{4\left(1 + 2ax^2\right)} \tag{31}$$

The metric functions $e^{2\lambda}$ and $e^{2\nu}$ can be written as

$$e^{2\lambda} = \frac{\left(1 + 2ax\right)}{\left(1 + ax\right)\left(1 - bx\right)} \tag{32}$$

$$e^{2\nu} = A^2c_2^2\left(-1 + bx\right)^{2E}\left(1 + ax\right)^{2F}\left(1 + 2ax\right)^{2G}\exp[2H(x)] \tag{33}$$

Figures 4, 5, 6, 7 and 8 represent the graphs of $p_r$, $\rho$, $\sigma^2$, $M(x)$ and $v_{sr}^2$ with eq.11, respectively and   a = 0.0169, b = 0.00454, α=1/3 with r=10 km . The metric for this model is

$$ds^2 = -A^2c_2^2\left(-1 + bx\right)^{2E}\left(1 + ax\right)^{2F}\left(1 + 2ax\right)^{2G}\exp[2H(x)]dt^2$$

$$+ \frac{\left(1 + 2ax\right)}{4xc\left(1 + ax\right)\left(1 - bx\right)}dx^2 + \frac{x}{c}(d\theta^2 + \sin^2\theta d\varphi^2) \tag{34}$$

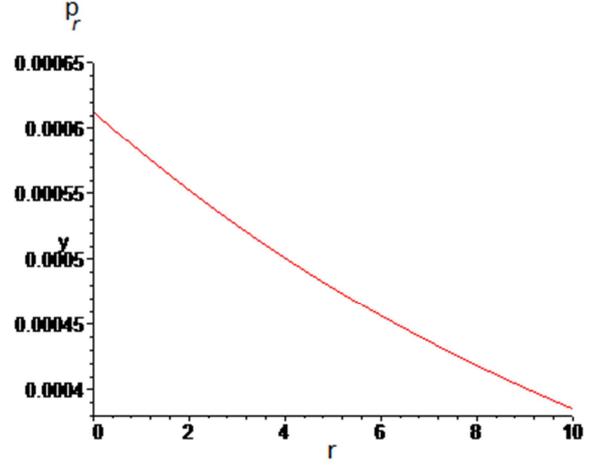

*Fig 4. Radial Pressure.*

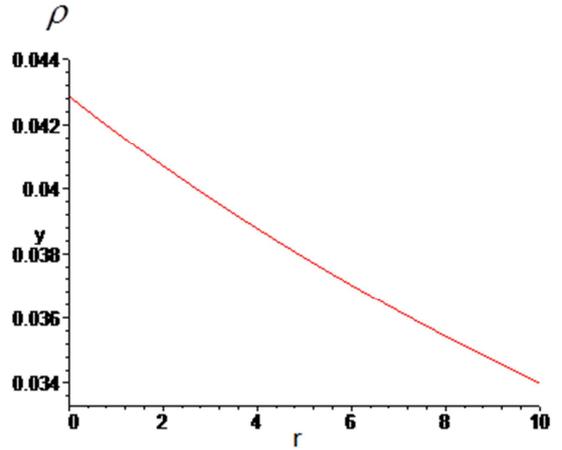

*Fig 5. Energy density*

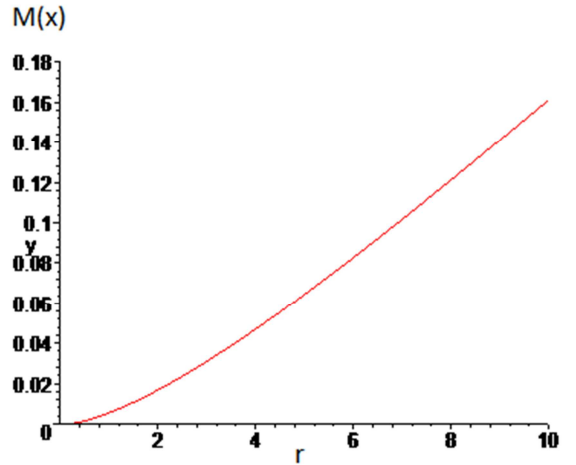

*Fig 6. Mass*

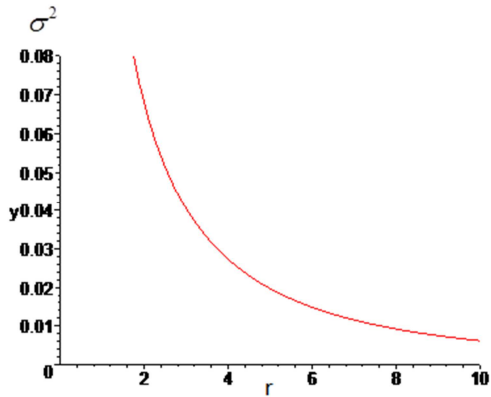

**Fig 7.** *Charge density*

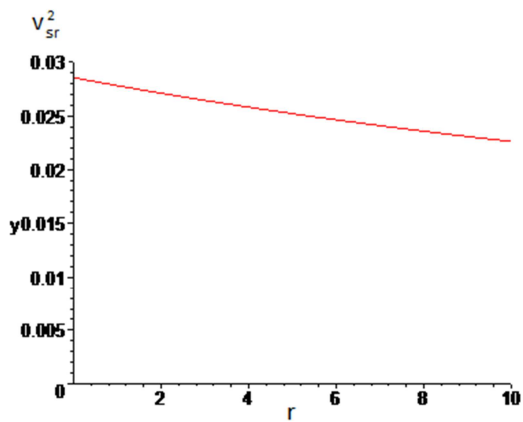

**Fig 8.** *Radial speed of sound*

## 4. Physical Properties of the New Solutions

In this section we discuss the physical properties that have to be satisfied by the realistic star [28]. With $E^2 = 0$, the gravitational potentials are regular at the origin since

$$e^{2\nu(0)} = A^2 c_1^2 e^{-\frac{2\alpha ac + 3\alpha bc}{2}}$$ and $e^{2\lambda(0)} = 1$ are constants and $\left(e^{2\lambda(r)}\right)' = \left(e^{2\nu(r)}\right)' = 0$ at r=0. In the centre $\rho(0) = \frac{3c(a+b)}{2}$

and $P_r = \frac{9\alpha c^2 (a+b)^2}{4}$ both are positive if $a > 0$ and $b > 0$.

For the case $E^2 = \frac{2c(1-Z)}{x}$ , $e^{2\lambda(0)} = 1$ , $e^{2\nu(0)} = A^2 c_2^2 e^{\frac{(2a-b)a\alpha}{2}}$ , in the origin $r = 0$ , $\left(e^{2\lambda(r)}\right)'_{r=0} = \left(e^{2\nu(r)}\right)'_{r=0} = 0$ . This shows that the potential gravitational is regular in the origin. In the centre $\rho(0) = 2c(a+b)$ , $P_r = \alpha c^2 (a+b)^2$ and the charge density presents a singularity. For both cases, the mass function is strictly increasing function, continuos and $M(x) = 0$ at r=0.

In fig.1 and fig. 4, the radial pressure is finite and decreasing for two studied cases. To maintain of causality,

the square of sound speed defined as $v_{sr}^2 = \frac{dp_r}{d\rho}$ should be within the limit $0 \leq v_{sr}^2 \leq 1$ in the interior of the star. In fig. 2 and 8 this condition is maintained inside the stellar interior. In fig. 5, that represent energy density for the case $E^2 \neq 0$ , we observe that is continuous, finite and monotonically decreasing function. In fig. 7, the charge density is singular at the origin, non-negative and decreases. In fig.3 and 6, the mass function is strictly increasing function, continuos and finite.

## 5. Conclusion

In this paper, we have generated new exact solutions to the Einstein-Maxwell system considering Tolman IV form for the gravitational potential Z and an equation of state that presents a quadratic relation between the energy density and the radial pressure. The new obtained models may be used to model relativistic stars in different astrophysical scenes. The relativistic solutions to the Einstein-Maxwell systems presented are physically reasonable. The charge density σ is singular at the origin for the case $E^2 \neq 0$ and the mass function is an increasing function, continuous and finite inside the stellar interior. The condition $0 \leq v_{sr}^2 \leq 1$ inside the stellar interior. The gravitational potentials are regular at the centre and well behaved.

The models presented in this article may be useful in the description of relativistic compact objects with charge, strange quark stars and configurations with anisotropic matter.